# ASTRONOMICAL AND HISTORICAL EVALUATION OF MOLNAR'S SOLUTION

Bradley E. Schaefer
Louisiana State University

## ABSTRACT
Previously, the most prominent explanation for the Star of Bethlehem was to identify one of many astronomical events in the sky as being the inspiration for the trip of the Magi. However, all the astronomical answers have detailed refutations. For example, comets, eclipses, and meteors were universally seen as very bad or evil omens, so they cannot be the Star. Also, there were no reported nova or supernova in 5 or 4 BC, because the Chinese reports explicitly identify both objects as being comets. And the claim that the Star was an earlier eruption of the 'recurrent nova' DO Aquilae is certainly wrong because DO Aquilae is not a recurrent novae (so it did not erupt in 5 BC) and it could not possibly have gotten to be bright enough to be visible to the unaided eye. The astronomical answers also have the general refutation that the spectacular astronomical events claimed to be the Star were all of no interest to the Magi. We do know exactly what is of interest to ancient astrologers and diviners, and they derive meaning and importance only from arcane patterns of the positions of the seven 'planets' as presented on a horoscope. There is no meaning to ancient astrologers for triple conjunctions or Venus/Jupiter occultations or any other spectacular astronomical event. The astrologers were looking down at their horoscopes and not upward to see any comet, nova, or supernova, and such were never placed onto a horoscope. With all this, all the astronomical answers for the Star are dead.

   In 1999, Michael Molnar put forth a completely new solution, where the Star originated as a report of a natal horoscope for 17 April 6 BC. This natal horoscope shows very impressive regal portents and points to Judea. It is very improbable that such a very rare planet configuration (averaging only once per millennium or longer) would coincide with the very restricted day of Jesus' birth (springtime in a year shortly before Herod's death in 4 BC), unless there is some causal connection. The Magi (as labeled by the gospel author) were astrologers, so they were only interested in horoscopes. The primary tool of the astrologer is the horoscope, and natal horoscopes tell exactly and only the date, country, character, and future for the birth of a child, and this is exactly the information that Matthew tells us that the Magi got from the Star. The astrological solution is further supported by the presence of an astrological technical term ('in the east' meaning what astronomers call the 'heliacal rise') in the Nativity narrative of Matthew. The astrological solution finally provides simple and natural explanations for many of the ways in which the Star operated. Importantly, Molnar has only identified the Star as originating from a particular natal horoscope, while making no statement about the nature or historicity of any of the other elements of the story in Matthew.  So for example, the Magi might be astrologers as idealized by the Greeks of the time, or they might have been non-existent (invented by a latter-day Greek seeking omens for the birth of a great king), and they might have arrived on 17 April 6 BC or months later, and they might or might not have been three in numbers with given names.  But for the Star of Bethlehem alone, Molnar's astrological solution is convincing.

## FOUR CLASSES OF ANSWERS

People have long wondered as to the nature of the Star of Bethlehem.  All explanations can be categorized as a 'pious fable', some 'astronomical event', a 'miracle', or an 'astrological horoscope'.

A 'pious fable' is some story invented with no basis in fact or evidence, imagined to satisfy or illustrate the inventor's faith.  In the case of the Star, the pious storymaker was presumably a devout Christian, likely recently converted, sometime shortly before the Gospel of Matthew was written.  With this, the entire story of the Star, the Magi, and their visit to Herod & Bethlehem is completely imaginary, with the Magi and the Star being nonexistent.  Indeed, we already confidently know that the Star story has attracted many pious fables; including the number and names of the Magi as well as the story of the Little Drummer Boy.  This Star explanation is favored by Bible-skeptics who seek to find factual or historical errors in anything related to the Bible.

The Star as an astronomical event is a real happening up in the sky.  The idea is that some presumably-spectacular celestial light show was interpreted as the Star and motivated the Magi to make the trip to Judea.  This class of Star-explanations has many claimed answers, which now includes most of the phenomena visible to the unaided eye (plus some too faint to be visible by anyone), both real and imagined.  This class of Star answer is favored by astronomers, simply because it is all they can discuss.  Thus, the common Christmas shows at planetariums worldwide only talk about the astronomical answers because they can uniquely illustrate the proposals and because laypersons can readily understand the claims.  This class of answers is also favored by Christian apologists, in an attempt to provide a rational basis for their faith.  This class of answers was also the nearly-exclusive concern of academic discussions of the Star.

The Star could be a 'miracle', where the deity worked outside conventional physics and astronomy so as to create a light in the sky that was seen and reported as the Star.  Within the framework of Christianity, this is a very plausible answer, because many miracles were occurring at nearly the same time and place (e.g., the Virgin Birth), while the importance of the event makes for a nontrivial reason to invoke a miracle.  While I do not have any formal survey, I expect that this 'miracle' answer is the predominant belief of lay Christians worldwide.

The 'astrological horoscope' answer is the idea that the Star is a report ultimately based on a configuration of planets resulting in a regal horoscope related to Christ's birth.  Critically, this answer does not require the Biblical-literalist interpretation of the story in Matthew, so the astrological horoscope explanation can be correct whether or not the Magi were Hellenized astrologers or whether they even existed.  Up until 1999, scholars had essentially ignored the astrological aspects of the Nativity story in Matthew.  Before 1999, Michael Molnar had been systematically applying ancient astrology knowledge to the historical study of ancient kings around the Mediterranean and the Middle East.  With this strong basis for knowing the real practices of ancient astrologers, Molnar (1999) provided the first look at the Nativity story from an astrological perspective, and he has identified a convincing case that the Star was a regal horoscope for 17 April 6 BC. Our conference in Groningen was explicitly an interdisciplinary and nondenominational study and popularization of Molnar's answer.

## ASTRONOMICAL ANSWERS

The astronomical class of answers all feature some specific celestial event that is perceived to be sufficiently spectacular enough that it might have attracted the attention of the Magi, sending them to Judea. This class of answers is possible only because modern astronomers can know the details of the ancient skies with remarkable accuracy. That is, the positions and characteristics of the Sun, Moon, planets, comets, and meteor showers are confidently known to high accuracy even many millennia ago, while historical records (like from China) alert us to various transient events. With this sure knowledge, many astronomers have cast backwards in time to the first decade BC, looking for something spectacular enough to be considered as the Star.

*Modern* astronomers and historians have found many celestial events that they consider to be spectacular enough to be the Star. See Table 1 for a partial listing of the claimed Stars. Some of the items in the table exist only as speculation. Most previous scholarly discussions, most popular books, and all Christmas shows at planetariums merely feature some subset of this list, often with some one possibility selected out as being best.

The long list of possibilities is striking in that there is no real way to distinguish amongst the many possibilities. That is, there is no significant evidential basis for selecting any one possibility as being better or worse than other events on the list. This is illustrated by the fact that popular books and scholarly articles all select out *different* events from this list as their favored Star.

Critically, we realize that only zero or one of the events listed in Table 1 can be the real Star, and that means that 12 or 13 of the claims must be wrong. With almost all of these claims being certainly wrong and with no way to distinguish amongst the astronomical answers, we further realize that there is no possibility to get any useful level of confidence for any such astronomical answer.

Why are there so many spectacular astronomical events, all crowded into the critical decade? Well, the reason is that *all* decades are crowded with spectacular astronomical events. Schaefer (1989; 2003) has presented detailed statistics, concluding that every decade has an average of 12 spectacular astronomical events, and many examples are presented for past decades. To make this point for a case that we all have as a living memory, I have compiled a partial list of astronomical spectacular events from the 2000-2009 decade that were considered spectacular enough to attract media attention (see Table 2). Two millennia from now, historians could look back and select any of these events as being important for the people of the time. With this plethora of events to choose from, it becomes like a Rorschach Test for the Star identifier. More generally, the sky is always filled with spectacular events, so willful prophets and historians can always select something useful to blame/warn/alert their flocks. With every decade having many real (and imagined) astronomical spectacular events, it is no surprise that the 12 BC to 2 BC time interval has more than a dozen.

The general class of astronomical answers has always had problems with a variety of aspects of the Nativity narrative. In particular, the astronomical answers have never provided answer to the three questions (1) "Why did the Magi see the Star in the east and then travel west?", (2) "Why did they travel first to Jerusalem instead of Bethlehem?", and (3) "Why did no one in Jerusalem see the spectacular Star?".

Many astronomical answers have strong refutations. Let me first highlight the refutations of a recent widely spread claim that appeared in a book by M. Kidger (1999). The idea is that the Star was a recurrent nova that appeared in 5 BC, as reported by the Chinese, with the particular nova erupting again as Nova Aquilae 1925 (designated DO Aql). One simple refutation is that the Chinese reported the 5 BC event as being a "broom star" (with the 4 BC event being labeled as a "fuzzy star") so it was certainly a comet and not a nova of any type. Two other complete refutations require some modern astronomical knowledge of DO Aql. For this, I happen to be the world's leading expert on DO Aql, the identity of novae that are really recurrent novae, and I have even made DO Aql into the prototype of my 'F class' slow novae (Schaefer 2013; Strope, Schaefer, & Henden 2010; Pagnotta & Schaefer 2014). From this work, we know with very high confidence that DO Aql is a slow nova with a low-mass white dwarf that recurs only with a very long cycle of longer than a million years, so it certainly did not have an eruption in 5 BC (Schaefer 2013). Further, DO Aql erupted to only 8.5 magnitude in 1925 (barely visible in binoculars) and any prior eruption could not have gotten to the required brightness (claimed to be V=0 magnitude by Kidger) without using supernova-like energies that would have completely destroyed the system.

Many of the other astronomical answers have refutations that are equally convincing. Let me list here a brief version of these refutations: (1) The first astronomical answer is that of Kepler, where he thought that a triple conjunction spawned a supernova. But modern astronomers know that there is no such connection between apparent planet positions and distance supernovae. (2) The Venus-Jupiter occultation might have been very rare and spectacular for a modern astronomer, but it occurred in 2 BC, while Herod certainly died in early 4 BC. (3) Supernovae and novae have been claimed to appear in 5 BC or 4 BC as based on the Chinese reports, but both of these are certainly not supernovae or novae because the Chinese reports call the transients as a "broom star" and a "fuzzy star". (4) The Star is frequently attributed to being a comet, either one of the Chinese comets or the Halley's Comet return in 12 BC. But comets are universally feared as evil omens (Schaefer 1997) and thus have no chance of being considered as the Star. (5) Meteors are occasionally pointed to as being the Star, but meteors are also universally feared (Schaefer 1998) and thus have no chance of inspiring the Star. (6) Lunar and solar eclipses are also universally feared (Schaefer 1992; Schaefer 1994). (7) The possibility of a 'hypernova' (i.e., an extreme supernova creating a Gamma-Ray Burst) in the Andromeda Galaxy is refuted because such events cannot become bright enough to be recognized by the Magi.

Molnar (1999) has a general refutation of most astronomical answers. His point is that the only people whose impressions of the Star matter are the Magi, and the Magi were taken by the Greeks of the time to be *astrologers*, not astronomers, and all the spectacular astronomical events are meaningless and un-noticed by ancient astrologers.

In particular, astrologers do not study the sky in any modern sense, and so they are not likely to have even seen any supernova, nova, or comet.  Astrologers have no place or symbol on their horoscopes for comets, occultations, nova, supernova, hypernova, Uranus, meteors, or any other spectacular astronomical event that would impress modern astronomers.  Ancient astrologers have no interpretation for triple conjunctions, occultations, meteors, or any other astronomical spectaculars.  Molnar's point is that we know well what would interest ancient astrologers and what would be useful and interpreted by ancient astrologers (e.g., from Ptolemy's *Tetrabiblos*), and all of the spectacular astronomical  events are completely irrelevant to the Magi.  This holds whether the Magi were the historical personages or were the Greek idealization of the Magi as astrologers.  With all of the astronomical answers being unrecognized by ancient astrologers, we know that all the astronomical answers are wrong.

In summary, strong arguments have been made that astronomical answers are wrong specifically and also in general.  Nevertheless, it is likely that one or another astronomical answer will be highlighted in future works by astronomers (because specific and evocative models are made), by planetariums (because astronomical Stars make for simple answers in a traditional show), by future apologists (because some naturalistic explanation is needed by their philosophy), and by lay people (because their parents told them these claims when they were children).  As scholars, we can read the above and realize that the astronomical answers are dead, but we should also realize that sociological forces and inertia make for a long time before the death notice is widely acknowledged.

## THE ASTROLOGICAL SOLUTION

Molnar (1999) presented the astrological solution for the Star.  This book was the culmination of a decade long series of scholarly studies of ancient astrology as applied to a wide variety of kings throughout the eastern Mediterranean region.  Thus Molnar has papers on the horoscopes of Julius Caesar, Caesar Augustus, Domitian, Mithridates, and many more.  These articles all treat astrology as a real historical force, and they all take ancient astrology from the original sources (e.g., Ptolemy, Firmicus).  Molnar's articles all appeared in an obscure (but top quality) numismatic journal called *The Celator* (Molnar 1993a; 1993b; 1993c; 1993d; 1994a; 1994b; 1995; 1996; 1997), and are now essentially unavailable.  I had been following Molnar's work in *The Celator* for many years before his book on the Star, and I had already been impressed both by the quality of his work and by the need to reject my old bias against astrology for historical applications.  Indeed, I suspect that the reason Molnar's work appeared only in a small-circulation niche journal was because of the main-stream contempt of astrology.

A substantial reason for why no one had previously come up with the astrological solution is simply that modern astronomers and scholars of many types have a repugnance to astrology.  This contempt has made for astrology and its effects being ignored and shunned by most scholars.  Those few scholars to seriously study astrology have produced small amounts of excellent work over the past century, but the work has always been disregarded and sidelined by most.  This contempt is completely wrong for *historians*.  Astrology has zero effectiveness as a predictive science, but this is irrelevant to the fact that astrology has been a real force acting on people and history.  The general historical community should remove astrology from its blindspot.

Molnar's astrological solution starts by considering what would be important to an ancient astrologer. (Critically, this might be either to the historical Magi of the ancient Near East or to the idealized Magi as pictured by Greeks around the time of the gospel writer.) The primary tool and instrument of astrologers is the horoscope, a schematic position of the planets (including the Sun and the Moon) within the zodiac and within the sky for a given time and place. A natal horoscope is for the time and place of the birth of a child, with the horoscope telling the character and future of the child. Matthew reports that the Star tells the astrologers about the date, place, character, and future of the birth of a child. That is, the Star tells the astrologers about the birth of a very great king to be born in Judea on some date, and that is exactly what a natal horoscope gives. It seems to be past coincidence that the Star in Matthew tells the ancient astrologers exactly what a natal horoscope tells to ancient astrologers. Thus, Molnar concludes that the Star was actually a report of a natal horoscope.

So suddenly, we have a good rationale for how the Star operates. Prior astronomical answers always had trouble explaining why the Magi went to Judea and why they were expecting the birth of a great king, Prior pious fable answers and miracle answers could include such details for no good reason. Now, Molnar's astrological solution provides a natural explanation of how the Magi knew that there would have been the birth of a very great king on a particular day in Judea.

Molnar goes further, searching through time for a regal horoscope pointing to Judea. From his earlier work, as based on Firmicus, Ptolemy, Antigonus, Valens, and Manilius, he knows what points to a regal horoscope, and what will emphasize the regal aspects. He also knows what will point to Judea, with the closest geographical-astrology list in time being Ptolemy, where the sign Aries is associated with Judea. (Molnar has two further arguments connecting Aries and Judea in the first century AD, based on coins of Antioch as well as based on the horoscope of Nero, but I judge these to be weaker than the straight statement in the *Tetrabiblos*.) With this, Molnar recognized that the date 17 April 6 BC has an impressive regal horoscope indicating the birth of a very great king in Judea. The regal omens pile on each other emphasizing the greatness of the king, while most of the key planets are in Aries hence pointing to Judea. Table 3 gives a list of the key aspects of the 17 April 6 BC natal horoscope.

Molnar's astrological solution provides easy and natural explanations for the three questions that are so hard for the astronomical answers. (1) Why did the Magi see the Star 'in the east' and then go west? The term "in the east" is an astrological technical phrase, meaning what we now call as a heliacal rising. Heliacal risings happen on one day about once a year for the planets, and would be visible, if anyone looked, low on the eastern horizon in the middle of dawn. Old astrologers would not have looked in the dawn skies, but on their horoscopes and seen the heliacal rising. Such a heliacal rising event does have significance for ancient astrology, and it would be part of a larger pattern. This pattern would point to the province of Judea. So the meaning of the passage in Matthew is that they have seen the Star at its helical rising and then gone to the capital of Judea. (2) The astrological solution also provides a fast and easy explanation for why the Magi first went to Jerusalem, instead of Bethlehem. The reason is that the ancient astrology could only specify the country or province. The regal horoscope pointed to Judea, so the Magi went to the capital of Judea and asked around, and this is exactly what Matthew says they do. For an analogy, suppose that some soothsayer in Germany had wanted to attend our conference in Groningen but only knew from the omens that it was in the Netherlands, then they would travel to the capital of the Netherlands and ask around until someone tells them to go to Groningen. (3) Why did no one in Jerusalem see the Star? Many spectacular astronomical events would be hard to miss. But the Jews in Jerusalem had little knowledge or practice of astrology, and the special patterns of the planets can only be recognized through the eyes of an astrologer looking at the positions on a horoscope. Important parts of the pattern can include planets close to the Sun or in daytime (hence invisible outside a horoscope) or planets separated in trines, integral multiples of 120° (something that the uninitiated would never spot). So of course no one in Jerusalem would have seen the arcane astrological pattern up in the sky.

Let me make an estimate of the frequency of such regal horoscopes pointing to Judea. Jupiter is at its heliacal rise (12° from the Sun) once each year, and this occurs in Aries once every twelve years on average. The Sun will not also be in Aries around 12/30=40% of the time, so the long-term average is once every 20 years with both the Sun and the heliacally-rising Jupiter in Aries. For these days, the Moon will be within 1.5 days of conjunction with the heliacally rising Jupiter around 10% of the time, making the average rate for such events as once every ~200 years. With the Sun in the western 18/30=60% of Aries, Venus will be at exaltation in Pisces for roughly one-fifth of the occasions, making for a long-term average of once per ~1000 years. For the aspects of the 17 April 6 BC horoscope, Saturn must also be in Aries (to be in trine with Jupiter and in attendance before the Sun) and this happens one time out twelve, for a rate of once every ~12,000 years. If we further require that the planets Mars and Mercury do not have negative portents (e.g., being in quartile with the Sun), we get a frequency of once every >>12,000 years. Similar conclusions have been reached by Dworetsky & Fosset (1998). Molnar (1999) points out that this idealized calculation has the substantial problem that it calculates the frequency for only one configuration (Jupiter/Sun/Saturn in Aries, the Moon in conjunction with Jupiter at its heliacal rise, and Venus in Pisces), whereas other very-rare configurations might also have produced a regal horoscope for Judea of comparable power. For example, with a weakening due to Saturn not being in attendance to the Sun, Saturn might have been in Leo or Sagittarius and still been in the trine, and this is twice as frequent as the case calculated. Unfortunately, it is difficult to know the number and nature of the planet configurations that would be adequate to inspire a report of a regal horoscope of enough power. Nevertheless, it is clear that the astrological requirements for a horoscope indicating the birth of a very great king in Judea are realized only once per many centuries. Historically, the horoscope for the Star is a very rare event, happening only one day out of something like a millennium or so. This is the reason that the Magi were inspired to travel to the west, or at least the reason that inspired a latter-day Greek omen-seeker to recognize the Star horoscope.

At our conference and soon after, I was asked about the accuracy of the calculations and the date of the heliacal rise of Jupiter. For such issues, I do have an extensive and definitive knowledge, both from theory and observation (e.g., Schaefer 1987; 1993; Doggett & Schaefer 1994). All the planets (i.e., everything in Molnar's horoscope) are tightly clusters around the Sun, as shown in Figure 2. At dawn on the morning, Jupiter was 12.4° from the Sun, with this changing at the rate of one degree per day. The astronomers and astrologers of the time likely could calculate these positions to an accuracy of around one degree or perhaps a bit worse. Ptolemy gives the criterion for Jupiter's heliacal rise to be a separation of 12° in ecliptic longitude from the Sun. So the Magi would have taken 17 April 6 BC as the date of heliacal rise, or possibly the day before or after. For the irrelevant further question of the date of the actual first visibility of Jupiter, the sharp angle of the ecliptic with respect to the horizon for this time of year makes for Jupiter having a low altitude (6.2° with no refraction) above the Sun, and this means that my best estimate for the date of first visibility is 17 or 18 April. This is for my accurate knowledge of the extinction coefficients in ancient Jerusalem in springtime (Schaefer 1990; 2001), where the typical variations in the haziness of the atmosphere makes for an uncertainty of a day or two. Even though my visibility dates coincide with the date from Molnar (to within the uncertainty of 1-2 days), the visibility date is irrelevant because the Magi (or some later Greek astrologer) would have used the dates from the astrological criterion.

At our conference, a variety of questions and problems were raised. Here are the answers to some of these: (1) The historical Magi were *not* Hellenized astrologers, as emphasized by several speakers, but this is not required for Molnar's solution. The reason is that the gospel writer was a Hellenized person, and the Greeks of the time idealized any eastern mystic as being what they called a "Magi", and the gospel description is going to be made from the perspective of the gospel writer with their terminology. So just because the gospel writer calls them Magi, does not mean that they correspond to what modern historians label as Magi. (2) Various speakers showed that a wide variety of geographical astrology lists do *not* point to Judea as being associated with Aries. Clearly the sign of Judea is not a constant. But most of the citations are for lists long before the time of Jesus. It is not relevant whether the first list, or any early list, does or does not have Judea identified with any sign. What matters is the list closest in time to the 6 BC event (or perhaps closest to the time of the gospel writer), and that is the list given in Ptolemy's *Tetrabiblos*. The *Tetrabiblos* explicitly gives Aries as being the sign of Judea. (3) Various people on the internet have characterized Molnar's solution as just being *one* of the aspects in Table 3, then belittled the solution. Well, no, because the regal horoscope involves many items, that *together* make for an impressive omen. It is wrong to pick out one aspect in isolation. (4) Once during the conference and many times on the internet, people have claimed that the lunar occultation of the heliacally rising Jupiter was non-existent, non-observable, or non-calculable by the Magi. Well, the occultation did occur for Jerusalem, it was unobservable because it happened during the daytime even though it was high in the sky, and the conjunction/occultation distinction could not have been calculated by the Magi. But this is all irrelevant because the ancient astrology makes no distinction between a conjunction and an occultation.

In summary, Molnar (1999) has provided a solution as to the nature of the Star of Bethlehem, and this solution has provided good explanations that had been impossible with prior answers.

# HISTORICITY OF THE *OTHER* PARTS OF THE NATIVITY STORY

Molnar's astrological solution tells us about the origin of the Star. But it does *not* tell us about the historicity of the other parts of the Nativity story in Matthew. To illustrate this, let me present two completely different scenarios, both where the Star in Matthew comes from the natal horoscope of 17 April 6 BC. (1) The first scenario has some group to the east of Jerusalem, whom the gospel writer calls 'Magi', recognizing the natal horoscope for 17 April 6 BC, travelling to Jerusalem and then Bethlehem, worshiping at the feet of the divine Christ child, and returning home by a different route. This scenario is as expected by pious Christians with a Biblical-literalist interpretation. (2) The second scenario has some unknown Greek person, likely around 70 AD or so, seeking a celestial omen for the birth of a great king, spotting the horoscope for 17 April 6 BC, recognizing this as an impressive regal horoscope, and inventing a pious fable about the Magi. As Molnar's long work in *The Celator* demonstrates, it is characteristic of the Greeks of the time to seek and give importance to celestial omens associated with the birth of all the great kings. It is inevitable that people would be casting back in time to the birth of Christ seeking omens in the sky, and they would have found the regal horoscope. Within this scenario, the Magi and their visit would have been invented, with no historicity other than that which the Gospel writer includes to give realism to the story. But the Star itself would still come from the 17 April 6 BC natal horoscope, with full historicity. This scenario is perfectly fine for people of all religions, historians, Bible-skeptics, as well as for pious Christians who are not Biblical-literalists.

These two scenarios are extremes, with a continuum of possibilities between (see Figure 2). The Star story could have been transmitted to the gospel writer by divine or angelic inspiration, or could have been transmitted from Mary (who was a direct witness) to early converts and hence to the gospel writer, or could have been told to the gospel writer by the latter-day Greek omen-seeker who recognized the 17 April 6 BC horoscope. The Magi might have made their trip as in Matthew, might have arrived late, or might have not existed. The Magi might have been the Magi known to historians or they might have been idealizations as Hellenized astrologers as commonly viewed by Greeks at the time of the gospel writer. (The Magi might have arrived at the manger on 17 April 6 BC, or they might have discovered the horoscope only retrospectively and arrived in Bethlehem months after Jesus' birth.) The historicity of the Star as originating with the natal horoscope of 17 April 6 BC says nothing about the historicity of the rest of the Bible, nor indeed does it say anything about the divinity of Jesus.

A very wide range of possibilities exist within the astrological solution, all with the Star originating as a historical report of the 17 April 6 BC natal horoscope. This range of possibilities easily fits around the preconceptions of biblical-literalist Christians, non-biblical-literalist Christians, Christians of all denominations, devout members of all religions, agnostics, Bible-skeptics, and atheists.

## OVERVIEW & CONCLUSIONS

The astronomical answers are dead, both by specific refutations of individual claims, and by the general realization that any spectacular astronomical event is meaningless for ancient astrologers.

The pious fable answer might still be correct, especially as an explanation for the Magi part of the Nativity story. Independent of your predisposition to believe or not-believe this possibility, it is rather hard to prove or disprove the pious fable hypothesis. Nevertheless, a story invented independent of astrology is rather unlikely to have included an astrological technical term ('in the east' for 'heliacal rise') or to have included exactly the information that astrologers would derive from a natal horoscope (time, place, country, character, and future of a child's birth).  A story invented independently from knowledge of the 17 April 6 BC horoscope is very unlikely because it cannot explain the coincidence of such a rare and powerful regal horoscope in the spring of a year soon before the death of Herod.  While a pious fable might be possible, scholars reject this as having no evidence in its favor, and as having explained none of the details or 'coincidences'.

      A miracle answer is always possible.  One speaker at the conference argued that the events in Matthews were of low probability of occurring simultaneously, but this is exactly the requirement for it to be a miracle.  If you believe in miracles, then you would nevertheless shy away from having the Star be a miracle because that would require the deity to be deceitful for having set up the miracle while also having such perfect evidence for the 6 BC horoscope as being the Star.  So pious Christians join with Christian apologists, Bible-as-history Christians, and non-Christians in rejecting the miracle answer.  For historians and other scholars, the key point is that the possibility of a miracle has zero positive evidence.

      In sharp contrast to the other possibilities, Molnar's astrological solution has many strong arguments going for it (see the summary in Table 4).  Finally, we have simple and natural explanations for the operation of the Star.  And the indicated birth date is just what we independently expect (i.e., in the springtime of a year shortly before Herod's death in 4 BC).  The astrological solution is dictated because a horoscope is the only thing that would interest the Magi, and a natal horoscope tells the ancient astrologers exactly what the Star tells the Magi in the second chapter of Matthew.

      Up until the time of Kepler, the miracle answer was the default idea to explain the Star.  Then, for the next four centuries, scholarly conclusions on the nature of the Star have settled on one of the various astronomical answers.  Now, with the new millennium, Molnar (1999) has finally provided a convincing answer, as convincing as any such historical question can allow.  Our conference in Groningen has been a popularization and a celebration of the new paradigm that has swept the field.  Molnar's astrological solution is now the leading and default explanation for the Star.  So with reasonable confidence, we now know that the Star of Bethlehem originated as a report of the natal horoscope for 17 April 6 BC.

## TABLE 1. Claimed Astronomical 'Stars'

- ★ 12 BC      Halley's Comet
- ★ 9 & 6 BC      Uranus passing by Saturn and Venus
- ★ 8 BC      Jupiter/Mars/Saturn conjunction
- ★ 7 BC      Jupiter/Saturn triple conjunction
- ★ 6 BC      Lunar occultations of Jupiter and Saturn
- ★ 5 BC      Stationary points of Jupiter
- ★ 5 BC      Hypernova in the Andromeda Galaxy
- ★ 5 BC      Chinese comet
- ★ 5 BC      Chinese nova or recurrent nova (DO Aql)
- ★ 4 BC      'Supernova'
- ★ 4 BC      Chinese 'nova'
- ★ 4 & 2 BC      *Two* Supernovae
- ★ 2 BC      Venus/Jupiter occultation

## TABLE 2. Spectacular Astronomical Events for the 2000-2009 Decade

- ★ 2000 & 2001      Great Leonid meteor storms
- ★ 2000, 2004      'End-of-World' planetary conjunctions
- ★ 2001, 2002, 2003, 2005, 2006, 2008, 2009
       Seven *total* solar eclipses
- ★ 2007      Great Comet McNaught
- ★ 2000 & 2001      Millennium change
- ★ 2000, 2000, 2001, 2003, 2003, 2004, 2004, 2007, 2007, 2008
       Ten *total* lunar eclipses
- ★ 2003      *Total* Lunar & Solar eclipses during Ramadan
- ★ 2002, …      'Super-Mars' and its echoes

## TABLE 3. Regal Aspects of the 17 April 6 BC Horoscope

- ★ Sun, Jupiter, and Saturn are in trine
- ★ Sun at Exaltation
- ★ Venus at Exaltation
- ★ Jupiter in Aries
- ★ Jupiter at heliacal rise with the Moon
- ★ Jupiter and Saturn in attendance before Sun

## TABLE 4. Molnar's Astrological Solution is Strong

- ★ Natal horoscopes are the tools of ancient astrologers, they tell only the birth/date/place/future of a child, with this being exactly the information that the Star tells the ancient astrologers in the Nativity story.
- ★ The Matthew account of the Star contains an astrological technical term, with 'in the east' meaning the 'heliacal rise'.
- ★ The horoscope for 17 April 6 BC has impressive regal portents.
- ★ The horoscope for 17 April 6 BC points to Judea.
- ★ The natal horoscope points to a date that is perfect for the tight restrictions from other grounds, i.e., that the birth of Jesus is in the springtime and in a year shortly before Herod's death in 4 BC.
- ★ High-potency horoscopes are very rare, so it is very improbable that such would appear in the springtime of a year soon before the death of Herod, unless there is some causal connection.
- ★ Explains how the Star could point to the birth of a child and to Judea
- ★ Explains why Magi saw Star 'in the east', yet went to the west
- ★ Explains why Magi first went to capital of Judea, not to Bethlehem
- ★ Explains why no one in Jerusalem saw the Star

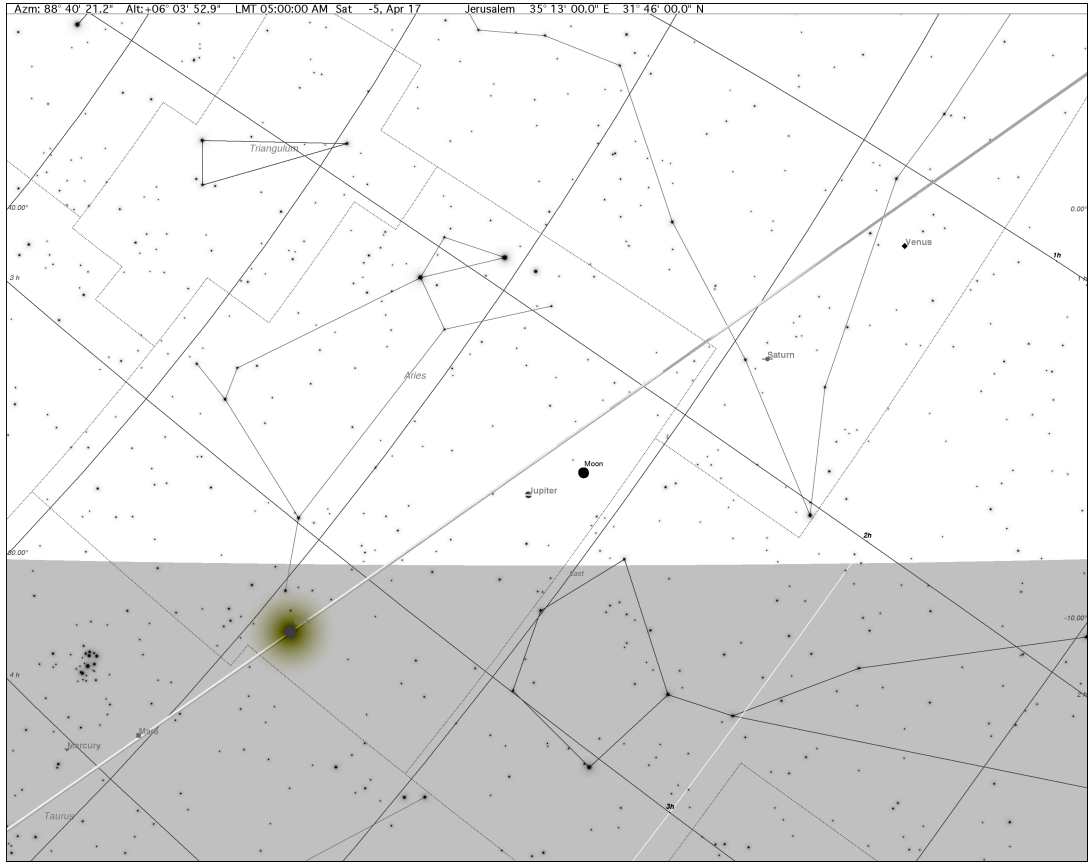

**Figure 1. The dawn sky on 17 April 6 BC from Jerusalem.**
Molnar's astrological solution points out that the Star of Bethlehem is the natal horoscope for 17 April 6 BC. Here are the positions of the Sun, Moon, planets, and stars for 03:00 UT for that morning as viewed from Jerusalem. The figure was made by the *Voyager* software, which I have checked against the latest highly-accurate ephemerides (including the definitive *JPL Horizons*) to have an accuracy of a few arc-seconds. The grey region on the bottom is the sky below the nominal horizon. The nearly diagonal line is the ecliptic. From lower left to upper right along the ecliptic, the planets are Mercury (just below the Pleiades), Mars, the Sun (in Aries), Jupiter (in Aries at its heliacal rise), the thin crescent Moon (fast approaching Jupiter for an unobservable occultation from Jerusalem in the middle of the day), Saturn (in the sign of Aries, but in the modern constellation of Pisces), and Venus (at exaltation in Pisces). At this time, Jupiter is 12.4° away from the Sun, with this changing by 1.0° throughout the day. Thus, by the astrological criterion for heliacal rising as given by Ptolemy, Jupiter was at its heliacal rise on 17 April 6 BC.

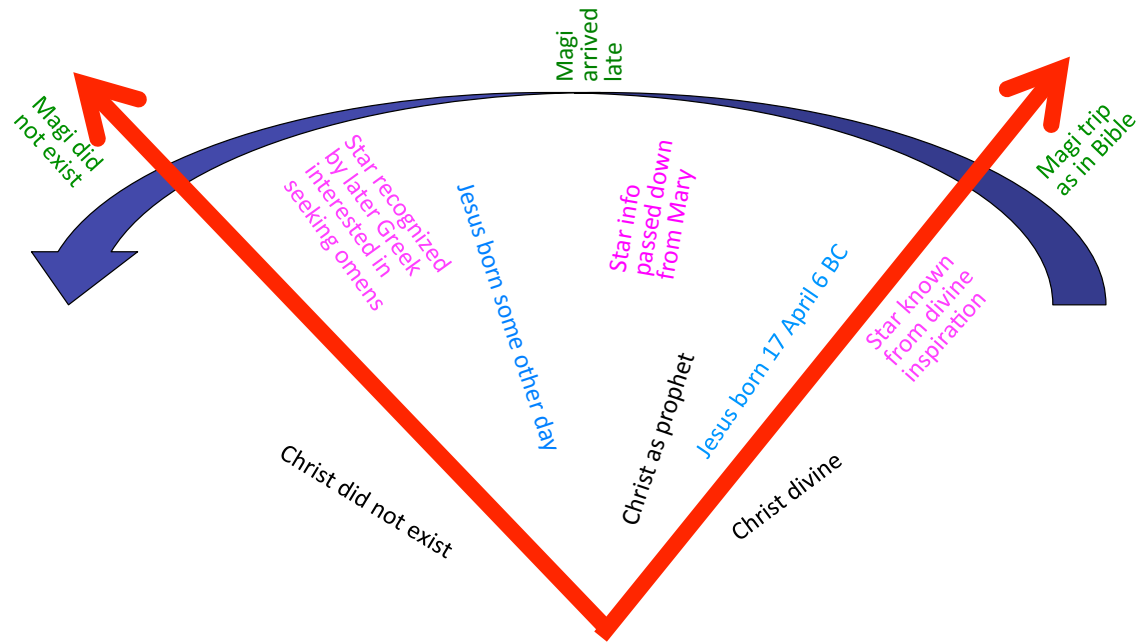

**Figure 2. Molnar has identified the Star, but the other elements remain unknown.** Molnar's astrological solution (the Star as the natal horoscope for 17 April 6 BC) does not say anything about the other elements of the biblical nativity story. There is a continuum of possibilities, all consistent with Molnar's discovery. The range extends from a Biblical-literalist view with everything exactly as stated in the Bible (with the arrow pointing to the right in the Figure), all the way to a historical solution where a latter-day Greek seeking omens for the birth of a great king spotted the 17 April 6 BC natal horoscope and invented the elements involving the trip of the Magi (with the arrow pointing to the left in the Figure). Thus, Molnar's solution is perfectly happy if the Magi were *not* Hellenized astrologers, if the Magi *were* Hellenized astrologers, if the Magi were three in number of any name, or if the Magi were non-existent. In all these cases, the real historicity of the Star remaining intact. Within Molnar's solution, Jesus' birth might or might not have been on 17 April 6 BC. Within Molnar's solution, the Magi might have arrived at Bethlehem on 17 April 6 BC, or they might have been months late. Within Molnar's solution, Christ might have been divine, a mere prophet, or non-existent. With these possibilities, the reader is free to turn the pointer to their favored scenario, all with the Star originating as a natal horoscope for 17 April 6 BC.